\documentclass[12pt]{article}
\usepackage{epsfig,a4,amsmath,amssymb}
\newcommand{\ee}       {\mbox{$\mathrm{e^+e^-}$} }

\newcommand{\roots}    {\mbox{$\sqrt{s}$} }
\newcommand {\qq}      {\rm{q}\overline{\rm{q}}}

\newcommand {\mt}      {m_{\mathrm{t}}}

\newcommand {\kZ}      {k_{\mathrm{Z}}}
\newcommand {\kg}      {k_{\gamma}}

\newcommand {\gevcc}   {{\rm GeV}/c^2}
\newcommand {\gevc}    {{\rm GeV}/c}
\begin{document}
\newpage
\begin{titlepage}
\pagestyle{empty}

\vskip 3.0 cm

\begin{center}
{\large EUROPEAN ORGANIZATION FOR NUCLEAR RESEARCH (CERN)}  
\vskip 0.3 cm
\end{center}
\begin{flushright}
{\tt CERN-EP/2002-xxx}\\
\today
\vskip 0.5 cm 
\end{flushright}

\vskip 2.0 cm

\begin{center}
\boldmath  
{\LARGE\bf Search for
Single Top Production\\in $\ee$ Collisions at $\roots$ up to 209~GeV}
\end{center}
\vskip 3.0 cm
\centerline{\Large
The ALEPH Collaboration$^{*)}$}
\vskip 3.0cm
\centerline{\large\bf Abstract}
\vskip 0.6 cm
{\small 
Single top production via the flavour changing neutral current reactions
 $\ee \to {\rm \bar{t} c , \bar{t} u}$ is searched for
 within the 214~pb$^{-1}$ of data collected by ALEPH at
centre-of-mass energies between 204 and 209~GeV. 
No deviation from the Standard Model expectation is observed and 
 upper limits on the single top production cross sections  are derived.
 The combination  with data  collected at 
lower centre-of-mass energies yields an upper limit on the branching ratio 
${\rm BR(t\to Zc)+BR(t\to Zu)}<14 \%$, for  
${\rm BR(t\to \gamma c)+BR(t\to \gamma u)}= 0$ and $\mt=174$~$\gevcc$. 

\vskip 1. cm
\begin{center}
{\it Submitted to Physics Letters B }
\vskip 1. cm
 *) See next pages for the list of authors. 
\end{center}
}

\end{titlepage}
\pagestyle{empty}
\newpage
\small
%
%
\newlength{\saveparskip}
\newlength{\savetextheight}
\newlength{\savetopmargin}
\newlength{\savetextwidth}
\newlength{\saveoddsidemargin}
\newlength{\savetopsep}
\setlength{\saveparskip}{\parskip}
\setlength{\savetextheight}{\textheight}
\setlength{\savetopmargin}{\topmargin}
\setlength{\savetextwidth}{\textwidth}
\setlength{\saveoddsidemargin}{\oddsidemargin}
\setlength{\savetopsep}{\topsep}
%
%
\setlength{\parskip}{0.0cm}
\setlength{\textheight}{25.0cm}
\setlength{\topmargin}{-1.5cm}
\setlength{\textwidth}{16 cm}
\setlength{\oddsidemargin}{-0.0cm}
\setlength{\topsep}{1mm}
\pretolerance=10000
\centerline{\large\bf The ALEPH Collaboration}
\footnotesize
\vspace{0.5cm}
{\raggedbottom
\begin{sloppypar}
\samepage\noindent
A.~Heister,
S.~Schael
\nopagebreak
\begin{center}
\parbox{15.5cm}{\sl\samepage
Physikalisches Institut das RWTH-Aachen, D-52056 Aachen, Germany}
\end{center}\end{sloppypar}
\vspace{2mm}
\begin{sloppypar}
\noindent
R.~Barate,
R.~Bruneli\`ere,
I.~De~Bonis,
D.~Decamp,
C.~Goy,
S.~Jezequel,
J.-P.~Lees,
F.~Martin,
E.~Merle,
\mbox{M.-N.~Minard},
B.~Pietrzyk,
B.~Trocm\'e
\nopagebreak
\begin{center}
\parbox{15.5cm}{\sl\samepage
Laboratoire de Physique des Particules (LAPP), IN$^{2}$P$^{3}$-CNRS,
F-74019 Annecy-le-Vieux Cedex, France}
\end{center}\end{sloppypar}
\vspace{2mm}
\begin{sloppypar}
\noindent
G.~Boix,$^{25}$
S.~Bravo,
M.P.~Casado,
M.~Chmeissani,
J.M.~Crespo,
E.~Fernandez,
M.~Fernandez-Bosman,
Ll.~Garrido,$^{15}$
E.~Graug\'{e}s,
J.~Lopez,
M.~Martinez,
G.~Merino,
A.~Pacheco,
D.~Paneque,
H.~Ruiz
\nopagebreak
\begin{center}
\parbox{15.5cm}{\sl\samepage
Institut de F\'{i}sica d'Altes Energies, Universitat Aut\`{o}noma
de Barcelona, E-08193 Bellaterra (Barcelona), Spain$^{7}$}
\end{center}\end{sloppypar}
\vspace{2mm}
\begin{sloppypar}
\noindent
A.~Colaleo,
D.~Creanza,
N.~De~Filippis,
M.~de~Palma,
G.~Iaselli,
G.~Maggi,
M.~Maggi,
S.~Nuzzo,
A.~Ranieri,
G.~Raso,$^{24}$
F.~Ruggieri,
G.~Selvaggi,
L.~Silvestris,
P.~Tempesta,
A.~Tricomi,$^{3}$
G.~Zito
\nopagebreak
\begin{center}
\parbox{15.5cm}{\sl\samepage
Dipartimento di Fisica, INFN Sezione di Bari, I-70126 Bari, Italy}
\end{center}\end{sloppypar}
\vspace{2mm}
\begin{sloppypar}
\noindent
X.~Huang,
J.~Lin,
Q. Ouyang,
T.~Wang,
Y.~Xie,
R.~Xu,
S.~Xue,
J.~Zhang,
L.~Zhang,
W.~Zhao
\nopagebreak
\begin{center}
\parbox{15.5cm}{\sl\samepage
Institute of High Energy Physics, Academia Sinica, Beijing, The People's
Republic of China$^{8}$}
\end{center}\end{sloppypar}
\vspace{2mm}
\begin{sloppypar}
\noindent
D.~Abbaneo,
P.~Azzurri,
T.~Barklow,$^{30}$
O.~Buchm\"uller,$^{30}$
M.~Cattaneo,
F.~Cerutti,
B.~Clerbaux,$^{23}$
H.~Drevermann,
R.W.~Forty,
M.~Frank,
F.~Gianotti,
T.C.~Greening,$^{26}$
J.B.~Hansen,
J.~Harvey,
D.E.~Hutchcroft,
P.~Janot,
B.~Jost,
M.~Kado,$^{2}$
P.~Mato,
A.~Moutoussi,
F.~Ranjard,
L.~Rolandi,
D.~Schlatter,
G.~Sguazzoni,
W.~Tejessy,
F.~Teubert,
A.~Valassi,
I.~Videau,
J.J.~Ward
\nopagebreak
\begin{center}
\parbox{15.5cm}{\sl\samepage
European Laboratory for Particle Physics (CERN), CH-1211 Geneva 23,
Switzerland}
\end{center}\end{sloppypar}
\vspace{2mm}
\begin{sloppypar}
\noindent
F.~Badaud,
S.~Dessagne,
A.~Falvard,$^{20}$
D.~Fayolle,
P.~Gay,
J.~Jousset,
B.~Michel,
S.~Monteil,
D.~Pallin,
J.M.~Pascolo,
P.~Perret
\nopagebreak
\begin{center}
\parbox{15.5cm}{\sl\samepage
Laboratoire de Physique Corpusculaire, Universit\'e Blaise Pascal,
IN$^{2}$P$^{3}$-CNRS, Clermont-Ferrand, F-63177 Aubi\`{e}re, France}
\end{center}\end{sloppypar}
\vspace{2mm}
\begin{sloppypar}
\noindent
J.D.~Hansen,
J.R.~Hansen,
P.H.~Hansen,
B.S.~Nilsson
\nopagebreak
\begin{center}
\parbox{15.5cm}{\sl\samepage
Niels Bohr Institute, 2100 Copenhagen, DK-Denmark$^{9}$}
\end{center}\end{sloppypar}
\vspace{2mm}
\begin{sloppypar}
\noindent
A.~Kyriakis,
C.~Markou,
E.~Simopoulou,
A.~Vayaki,
K.~Zachariadou
\nopagebreak
\begin{center}
\parbox{15.5cm}{\sl\samepage
Nuclear Research Center Demokritos (NRCD), GR-15310 Attiki, Greece}
\end{center}\end{sloppypar}
\vspace{2mm}
\begin{sloppypar}
\noindent
A.~Blondel,$^{12}$
\mbox{J.-C.~Brient},
F.~Machefert,
A.~Roug\'{e},
M.~Swynghedauw,
R.~Tanaka
\linebreak
H.~Videau
\nopagebreak
\begin{center}
\parbox{15.5cm}{\sl\samepage
Laoratoire Leprince-Ringuet, Ecole
Polytechnique, IN$^{2}$P$^{3}$-CNRS, \mbox{F-91128} Palaiseau Cedex, France}
\end{center}\end{sloppypar}
\vspace{2mm}
\begin{sloppypar}
\noindent
V.~Ciulli,
E.~Focardi,
G.~Parrini
\nopagebreak
\begin{center}
\parbox{15.5cm}{\sl\samepage
Dipartimento di Fisica, Universit\`a di Firenze, INFN Sezione di Firenze,
I-50125 Firenze, Italy}
\end{center}\end{sloppypar}
\vspace{2mm}
\begin{sloppypar}
\noindent
A.~Antonelli,
M.~Antonelli,
G.~Bencivenni,
F.~Bossi,
G.~Capon,
V.~Chiarella,
P.~Laurelli,
G.~Mannocchi,$^{5}$
G.P.~Murtas,
L.~Passalacqua
\nopagebreak
\begin{center}
\parbox{15.5cm}{\sl\samepage
Laboratori Nazionali dell'INFN (LNF-INFN), I-00044 Frascati, Italy}
\end{center}\end{sloppypar}
\vspace{2mm}
\begin{sloppypar}
\noindent
J.~Kennedy,
J.G.~Lynch,
P.~Negus,
V.~O'Shea,
A.S.~Thompson
\nopagebreak
\begin{center}
\parbox{15.5cm}{\sl\samepage
Department of Physics and Astronomy, University of Glasgow, Glasgow G12
8QQ,United Kingdom$^{10}$}
\end{center}\end{sloppypar}
\vspace{2mm}
\begin{sloppypar}
\noindent
S.~Wasserbaech
\nopagebreak
\begin{center}
\parbox{15.5cm}{\sl\samepage
Department of Physics, Haverford College, Haverford, PA 19041-1392, U.S.A.}
\end{center}\end{sloppypar}
\vspace{2mm}
\begin{sloppypar}
\noindent
R.~Cavanaugh,$^{4}$
S.~Dhamotharan,$^{21}$
C.~Geweniger,
P.~Hanke,
V.~Hepp,
E.E.~Kluge,
G.~Leibenguth,
A.~Putzer,
H.~Stenzel,
K.~Tittel,
M.~Wunsch$^{19}$
\nopagebreak
\begin{center}
\parbox{15.5cm}{\sl\samepage
Kirchhoff-Institut f\"ur Physik, Universit\"at Heidelberg, D-69120
Heidelberg, Germany$^{16}$}
\end{center}\end{sloppypar}
\vspace{2mm}
\begin{sloppypar}
\noindent
R.~Beuselinck,
W.~Cameron,
G.~Davies,
P.J.~Dornan,
M.~Girone,$^{1}$
R.D.~Hill,
N.~Marinelli,
J.~Nowell,
S.A.~Rutherford,
J.K.~Sedgbeer,
J.C.~Thompson,$^{14}$
R.~White
\nopagebreak
\begin{center}
\parbox{15.5cm}{\sl\samepage
Department of Physics, Imperial College, London SW7 2BZ,
United Kingdom$^{10}$}
\end{center}\end{sloppypar}
\vspace{2mm}
\begin{sloppypar}
\noindent
V.M.~Ghete,
P.~Girtler,
E.~Kneringer,
D.~Kuhn,
G.~Rudolph
\nopagebreak
\begin{center}
\parbox{15.5cm}{\sl\samepage
Institut f\"ur Experimentalphysik, Universit\"at Innsbruck, A-6020
Innsbruck, Austria$^{18}$}
\end{center}\end{sloppypar}
\vspace{2mm}
\begin{sloppypar}
\noindent
E.~Bouhova-Thacker,
C.K.~Bowdery,
D.P.~Clarke,
G.~Ellis,
A.J.~Finch,
F.~Foster,
G.~Hughes,
R.W.L.~Jones,
M.R.~Pearson,
N.A.~Robertson,
M.~Smizanska
\nopagebreak
\begin{center}
\parbox{15.5cm}{\sl\samepage
Department of Physics, University of Lancaster, Lancaster LA1 4YB,
United Kingdom$^{10}$}
\end{center}\end{sloppypar}
\vspace{2mm}
\begin{sloppypar}
\noindent
O.~van~der~Aa,
C.~Delaere,
V.~Lemaitre
\nopagebreak
\begin{center}
\parbox{15.5cm}{\sl\samepage
Institut de Physique Nucl\'eaire, D\'epartement de Physique, Universit\'e Catholique de Louvain, 1348 Louvain-la-Neuve, Belgium}
\end{center}\end{sloppypar}
\vspace{2mm}
\begin{sloppypar}
\noindent
U.~Blumenschein,
F.~H\"olldorfer,
K.~Jakobs,
F.~Kayser,
K.~Kleinknecht,
A.-S.~M\"uller,
G.~Quast,$^{6}$
B.~Renk,
H.-G.~Sander,
S.~Schmeling,
H.~Wachsmuth,
C.~Zeitnitz,
T.~Ziegler
\nopagebreak
\begin{center}
\parbox{15.5cm}{\sl\samepage
Institut f\"ur Physik, Universit\"at Mainz, D-55099 Mainz, Germany$^{16}$}
\end{center}\end{sloppypar}
\vspace{2mm}
\begin{sloppypar}
\noindent
A.~Bonissent,
P.~Coyle,
C.~Curtil,
A.~Ealet,
D.~Fouchez,
P.~Payre,
A.~Tilquin
\nopagebreak
\begin{center}
\parbox{15.5cm}{\sl\samepage
Centre de Physique des Particules de Marseille, Univ M\'editerran\'ee,
IN$^{2}$P$^{3}$-CNRS, F-13288 Marseille, France}
\end{center}\end{sloppypar}
\vspace{2mm}
\begin{sloppypar}
\noindent
F.~Ragusa
\nopagebreak
\begin{center}
\parbox{15.5cm}{\sl\samepage
Dipartimento di Fisica, Universit\`a di Milano e INFN Sezione di
Milano, I-20133 Milano, Italy.}
\end{center}\end{sloppypar}
\vspace{2mm}
\begin{sloppypar}
\noindent
A.~David,
H.~Dietl,
G.~Ganis,$^{27}$
K.~H\"uttmann,
G.~L\"utjens,
W.~M\"anner,
\mbox{H.-G.~Moser},
R.~Settles,
G.~Wolf
\nopagebreak
\begin{center}
\parbox{15.5cm}{\sl\samepage
Max-Planck-Institut f\"ur Physik, Werner-Heisenberg-Institut,
D-80805 M\"unchen, Germany\footnotemark[16]}
\end{center}\end{sloppypar}
\vspace{2mm}
\begin{sloppypar}
\noindent
J.~Boucrot,
O.~Callot,
M.~Davier,
L.~Duflot,
\mbox{J.-F.~Grivaz},
Ph.~Heusse,
A.~Jacholkowska,$^{32}$
L.~Serin,
\mbox{J.-J.~Veillet},
J.-B.~de~Vivie~de~R\'egie,$^{28}$
C.~Yuan
\nopagebreak
\begin{center}
\parbox{15.5cm}{\sl\samepage
Laboratoire de l'Acc\'el\'erateur Lin\'eaire, Universit\'e de Paris-Sud,
IN$^{2}$P$^{3}$-CNRS, F-91898 Orsay Cedex, France}
\end{center}\end{sloppypar}
\vspace{2mm}
\begin{sloppypar}
\noindent
G.~Bagliesi,
T.~Boccali,
L.~Fo\`a,
A.~Giammanco,
A.~Giassi,
F.~Ligabue,
A.~Messineo,
F.~Palla,
G.~Sanguinetti,
A.~Sciab\`a,
R.~Tenchini,$^{1}$
A.~Venturi,$^{1}$
P.G.~Verdini
\samepage
\begin{center}
\parbox{15.5cm}{\sl\samepage
Dipartimento di Fisica dell'Universit\`a, INFN Sezione di Pisa,
e Scuola Normale Superiore, I-56010 Pisa, Italy}
\end{center}\end{sloppypar}
\vspace{2mm}
\begin{sloppypar}
\noindent
O.~Awunor,
G.A.~Blair,
G.~Cowan,
A.~Garcia-Bellido,
M.G.~Green,
L.T.~Jones,
T.~Medcalf,
A.~Misiejuk,
J.A.~Strong,
P.~Teixeira-Dias
\nopagebreak
\begin{center}
\parbox{15.5cm}{\sl\samepage
Department of Physics, Royal Holloway \& Bedford New College,
University of London, Egham, Surrey TW20 OEX, United Kingdom$^{10}$}
\end{center}\end{sloppypar}
\vspace{2mm}
\begin{sloppypar}
\noindent
R.W.~Clifft,
T.R.~Edgecock,
P.R.~Norton,
I.R.~Tomalin
\nopagebreak
\begin{center}
\parbox{15.5cm}{\sl\samepage
Particle Physics Dept., Rutherford Appleton Laboratory,
Chilton, Didcot, Oxon OX11 OQX, United Kingdom$^{10}$}
\end{center}\end{sloppypar}
\vspace{2mm}
\begin{sloppypar}
\noindent
\mbox{B.~Bloch-Devaux},
D.~Boumediene,
P.~Colas,
B.~Fabbro,
E.~Lan\c{c}on,
\mbox{M.-C.~Lemaire},
E.~Locci,
P.~Perez,
J.~Rander,
B.~Tuchming,
B.~Vallage
\nopagebreak
\begin{center}
\parbox{15.5cm}{\sl\samepage
CEA, DAPNIA/Service de Physique des Particules,
CE-Saclay, F-91191 Gif-sur-Yvette Cedex, France$^{17}$}
\end{center}\end{sloppypar}
\vspace{2mm}
\begin{sloppypar}
\noindent
N.~Konstantinidis,
A.M.~Litke,
G.~Taylor
\nopagebreak
\begin{center}
\parbox{15.5cm}{\sl\samepage
Institute for Particle Physics, University of California at
Santa Cruz, Santa Cruz, CA 95064, USA$^{22}$}
\end{center}\end{sloppypar}
\vspace{2mm}
\begin{sloppypar}
\noindent
C.N.~Booth,
S.~Cartwright,
F.~Combley,$^{31}$
P.N.~Hodgson,
M.~Lehto,
L.F.~Thompson
\nopagebreak
\begin{center}
\parbox{15.5cm}{\sl\samepage
Department of Physics, University of Sheffield, Sheffield S3 7RH,
United Kingdom$^{10}$}
\end{center}\end{sloppypar}
\vspace{2mm}
\begin{sloppypar}
\noindent
A.~B\"ohrer,
S.~Brandt,
C.~Grupen,
J.~Hess,
A.~Ngac,
G.~Prange,
U.~Sieler
\nopagebreak
\begin{center}
\parbox{15.5cm}{\sl\samepage
Fachbereich Physik, Universit\"at Siegen, D-57068 Siegen, Germany$^{16}$}
\end{center}\end{sloppypar}
\vspace{2mm}
\begin{sloppypar}
\noindent
C.~Borean,
G.~Giannini
\nopagebreak
\begin{center}
\parbox{15.5cm}{\sl\samepage
Dipartimento di Fisica, Universit\`a di Trieste e INFN Sezione di Trieste,
I-34127 Trieste, Italy}
\end{center}\end{sloppypar}
\vspace{2mm}
\begin{sloppypar}
\noindent
H.~He,
J.~Putz,
J.~Rothberg
\nopagebreak
\begin{center}
\parbox{15.5cm}{\sl\samepage
Experimental Elementary Particle Physics, University of Washington, Seattle,
WA 98195 U.S.A.}
\end{center}\end{sloppypar}
\vspace{2mm}
\begin{sloppypar}
\noindent
S.R.~Armstrong,
K.~Berkelman,
K.~Cranmer,
D.P.S.~Ferguson,
Y.~Gao,$^{29}$
S.~Gonz\'{a}lez,
O.J.~Hayes,
H.~Hu,
S.~Jin,
J.~Kile,
P.A.~McNamara III,
J.~Nielsen,
Y.B.~Pan,
\mbox{J.H.~von~Wimmersperg-Toeller}, 
W.~Wiedenmann,
J.~Wu,
Sau~Lan~Wu,
X.~Wu,
G.~Zobernig
\nopagebreak
\begin{center}
\parbox{15.5cm}{\sl\samepage
Department of Physics, University of Wisconsin, Madison, WI 53706,
USA$^{11}$}
\end{center}\end{sloppypar}
\vspace{2mm}
\begin{sloppypar}
\noindent
G.~Dissertori
\nopagebreak
\begin{center}
\parbox{15.5cm}{\sl\samepage
Institute for Particle Physics, ETH H\"onggerberg, 8093 Z\"urich,
Switzerland.}
\end{center}\end{sloppypar}
}
\footnotetext[1]{Also at CERN, 1211 Geneva 23, Switzerland.}
\footnotetext[2]{Now at Fermilab, PO Box 500, MS 352, Batavia, IL 60510, USA}
\footnotetext[3]{Also at Dipartimento di Fisica di Catania and INFN Sezione di
 Catania, 95129 Catania, Italy.}
\footnotetext[4]{Now at University of Florida, Department of Physics, Gainesville, Florida 32611-8440, USA}
\footnotetext[5]{Also Istituto di Cosmo-Geofisica del C.N.R., Torino,
Italy.}
\footnotetext[6]{Now at Institut f\"ur Experimentelle Kernphysik, Universit\"at Karlsruhe, 76128 Karlsruhe, Germany.}
\footnotetext[7]{Supported by CICYT, Spain.}
\footnotetext[8]{Supported by the National Science Foundation of China.}
\footnotetext[9]{Supported by the Danish Natural Science Research Council.}
\footnotetext[10]{Supported by the UK Particle Physics and Astronomy Research
Council.}
\footnotetext[11]{Supported by the US Department of Energy, grant
DE-FG0295-ER40896.}
\footnotetext[12]{Now at Departement de Physique Corpusculaire, Universit\'e de
Gen\`eve, 1211 Gen\`eve 4, Switzerland.}
\footnotetext[13]{Supported by the Commission of the European Communities,
contract ERBFMBICT982874.}
\footnotetext[14]{Supported by the Leverhulme Trust.}
\footnotetext[15]{Permanent address: Universitat de Barcelona, 08208 Barcelona,
Spain.}
\footnotetext[16]{Supported by Bundesministerium f\"ur Bildung
und Forschung, Germany.}
\footnotetext[17]{Supported by the Direction des Sciences de la
Mati\`ere, C.E.A.}
\footnotetext[18]{Supported by the Austrian Ministry for Science and Transport.}
\footnotetext[19]{Now at SAP AG, 69185 Walldorf, Germany}
\footnotetext[20]{Now at Groupe d' Astroparticules de Montpellier, Universit\'e de Montpellier II, 34095 Montpellier, France.}
\footnotetext[21]{Now at BNP Paribas, 60325 Frankfurt am Mainz, Germany}
\footnotetext[22]{Supported by the US Department of Energy,
grant DE-FG03-92ER40689.}
\footnotetext[23]{Now at Institut Inter-universitaire des hautes Energies (IIHE), CP 230, Universit\'{e} Libre de Bruxelles, 1050 Bruxelles, Belgique}
\footnotetext[24]{Also at Dipartimento di Fisica e Tecnologie Relative, Universit\`a di Palermo, Palermo, Italy.}
\footnotetext[25]{Now at McKinsey and Compagny, Avenue Louis Casal 18, 1203 Geneva, Switzerland.}
\footnotetext[26]{Now at Honeywell, Phoenix AZ, U.S.A.}
\footnotetext[27]{Now at INFN Sezione di Roma II, Dipartimento di Fisica, Universit\`a di Roma Tor Vergata, 00133 Roma, Italy.}
\footnotetext[28]{Now at Centre de Physique des Particules de Marseille, Univ M\'editerran\'ee, F-13288 Marseille, France.}
\footnotetext[29]{Also at Department of Physics, Tsinghua University, Beijing, The People's Republic of China.}
\footnotetext[30]{Now at SLAC, Stanford, CA 94309, U.S.A.}
\footnotetext[31]{Deceased.}
\footnotetext[32]{Also at Groupe d' Astroparticules de Montpellier, Universit\'e de Montpellier II, 34095 Montpellier, France.}  
\setlength{\parskip}{\saveparskip}
\setlength{\textheight}{\savetextheight}
\setlength{\topmargin}{\savetopmargin}
\setlength{\textwidth}{\savetextwidth}
\setlength{\oddsidemargin}{\saveoddsidemargin}
\setlength{\topsep}{\savetopsep}
\normalsize
\newpage
\pagestyle{plain}
\setcounter{page}{1}

\setcounter{footnote}{0}
\setcounter{page}{1}
\textwidth=18.0cm
\normalsize
\pagestyle{plain}
\section{Introduction}
\label{intro}
Because of the large mass of the top quark~\cite{top_mass}, 
at LEP2 only single top production is possible. This could occur
via flavour changing neutral currents (FCNC) in the 
reactions\footnote{Throughout
this paper, the notation $\rm{\bar{t}q}$ is used for both $\rm{\bar{t} q}$ 
and $\rm{\bar{q} t}$ (q=c,u).}
\[
\ee \to {\rm \bar{t} c, \bar{t} u}\,. \label{FCNC} 
\]
In the Standard Model (SM) such a process is forbidden at tree level
and can only proceed via loops with cross sections 
$\lesssim 10^{-9}$~fb~\cite{huang}.
Extensions of the Standard Model could lead to enhancements of
FCNC single top production and to measurable effects 
as proposed, for example, in Refs.~[3-7].
It is customary to parametrise the FCNC transitions in terms of
anomalous vertices whose strengths are described by the parameters
$\kZ$ and $\kg$ for Z and $\gamma$ exchange, respectively,
using the formalism of Ref.~\cite{pecceia}.

 The results of searches for single top production are presented 
 in this letter with the data collected 
 by the ALEPH detector at LEP at centre-of-mass energies ranging from
 204 to 209~GeV, corresponding to an integrated luminosity of 214 pb$^{-1}$.
 Previous ALEPH results obtained with lower energy data are given
 in Ref.~\cite{tc1}.  
 The centre-of-mass energies and the integrated luminosities of the analysed
 data sample are listed in Table~\ref{tabellone}.
 
 The letter is organised as follows. The ALEPH detector is briefly reviewed in
 Section~2 together with the simulation samples used for the analysis.
  Section~3 is dedicated to the selection
  algorithm. In Section~4 the results of the searches are given, along with their
 interpretation within the theoretical framework of Ref.~\cite{pecceia}.
 The conclusions of the letter are given in Section~5.

\section{The ALEPH detector and the simulation samples}
\label{detector}

A thorough description of the ALEPH detector is presented in Ref.~\cite{Alnim},
and an account of its performance can be found in Ref.~\cite{Alperf}.
Only a brief overview is given here.

The tracking system consists of a silicon vertex detector (VDET),
a drift chamber (ITC) and a large time projection chamber (TPC),
immersed in a 1.5~T magnetic field produced by a 
superconducting solenoid.
The VDET consists of two concentric layers of double-sided
silicon microstrip detectors positioned at average radii of 6.5~cm and 
11.3~cm, covering 85\% and 69\% of the solid angle, respectively.
It is surrounded by the ITC, a multilayer axial-wire
cylindrical drift chamber.
The TPC 
provides up to 21 three-dimensional space coordinates and 338 samples 
of ionization loss  for tracks at radii between 30 and 180~cm.
Altogether, a transverse
momentum resolution of 
$\sigma(1/p_{\rm t})= 0.6 \times 10^{-3} \oplus 0.005/p_{\rm t}$
($p_{\rm t}$ in $\gevc$) is achieved.

The electromagnetic calorimeter (ECAL),  a lead/proportional-wire-chamber
sampling device of 22 radiation lengths, 
 surrounds the TPC and is contained inside the superconducting coil.
The energy resolution is  
$\sigma(E)/E=0.18/\sqrt{E} + 0.009$ ($E$ in GeV).

The magnetic field return yoke is a large iron structure
fully instrumented to form a hadron  calorimeter (HCAL), and
also serves as a muon filter.
The HCAL consists of 23 layers of streamer tubes
for  a total of 7.2 interaction lengths. 
The relative energy resolution  is 
$\sigma(E)/E=0.85/\sqrt{E}$  ($E$ in GeV).
 It is surrounded by two double layers of streamer tubes
 to improve the muon  identification. 

The luminosity monitors (LCAL and SICAL) extend the calorimetric coverage
down to polar angles of 34~mrad. 

The measurement from the tracking detectors and calorimeters are
combined into ``objects'', classified as electrons, muons, photons,
and charged and neutral hadrons, by means of the energy flow
algorithm described in~\cite{Alperf}.

The signal samples were simulated using the generator described in 
 Ref.~\cite{tc1}.
 A sample of 2000  events was produced for 
 each of the two final states  
 $\bar{\rm{t}}$c and $\bar{\rm{t}}$u at $\sqrt{s}$
 of  204,  206 and 209~GeV and for three values of the top mass
 (169, 174 and 179~$\gevcc$).

 The relevant SM background processes were simulated as follows.
 The  generators PYTHIA~\cite{tor} and  KORALZ~\cite{koralz} were
 both used for the $\qq$ simulation.
 The KORALW~\cite{koralw} generator was used to produce WW events, while
 the simulation of  We$\nu$ and ZZ events was based on PYTHIA.
 The size of the simulated samples typically corresponds to ten times 
 the integrated luminosity of the data.
 All background and signal samples were processed through the full detector
 simulation.

\section{Analysis}
\label{analysis}

 The processes  
 $\ee \to {\rm \bar{t} c,~ \bar{t} u} ~{\rm(\bar{t}\to \bar{ b} W^-)}$
 are characterised by a multijet final state with one b jet.
 The event properties depend significantly on the W decay mode.
 Two separate analyses have been designed for the 
  leptonic and the hadronic decays of the W.
 The selections follow closely
 those described in Ref.~\cite{tc1}.
 The minor changes in the hadronic selection arise from a re-optimization
 of the selection criteria according to the $\bar{N}_{95}$
 prescription~\cite{nbar95}, 
 to take into account the increase in the centre-of-mass energy
 and of the background level:
\begin{itemize}
\item The lower cut on the b-jet energy was loosened to ${\rm E(b jet)>50~GeV}$.
\item The lower cut on the W boost was loosened to $P_{\rm qq}/E_{\rm qq} > 0.5$.
\item The lower cut on the b-tag variable for the most b-like jet 
   cut has been tightened to ${\rm (b~tag)>6.4}$
\end{itemize}
 There were no changes in the leptonic selection.
\begin{figure}[t]
\mbox{\epsfig{file=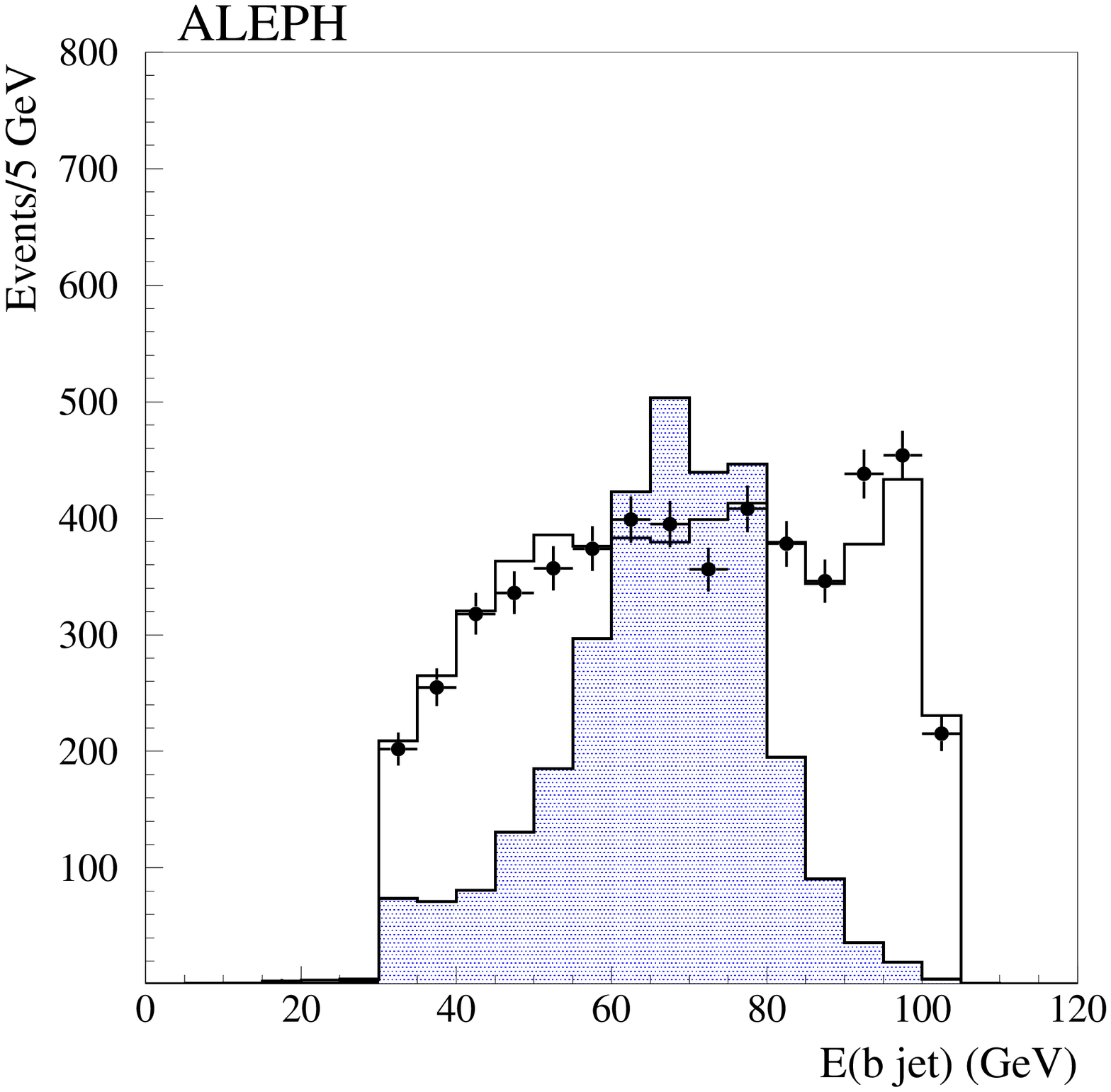,width=7.5cm,height=7.5cm} 
      \epsfig{file=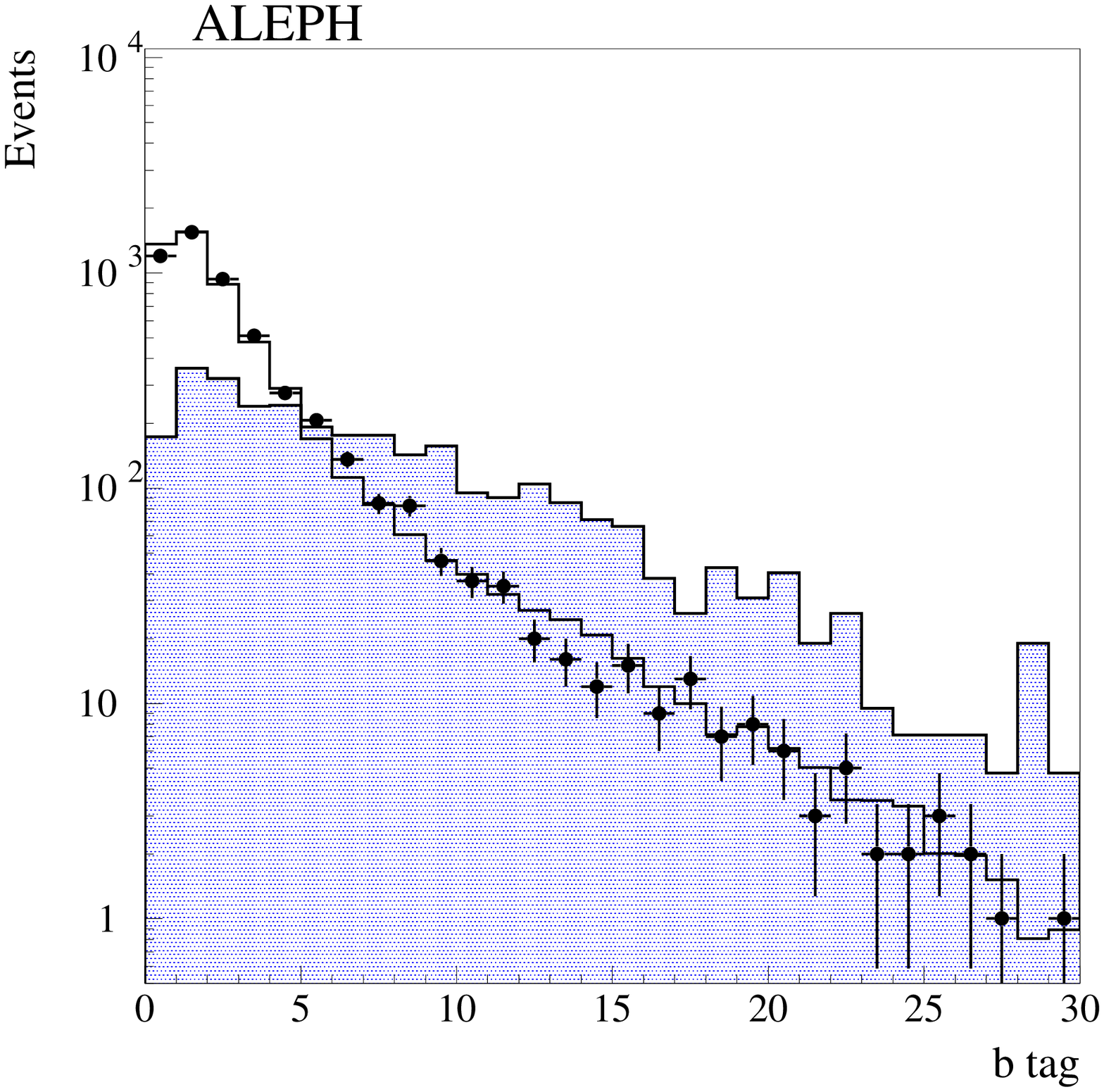,width=7.5cm,height=7.5cm} }
\caption
{\small Distributions for hadronic W decays for  
 the signal (shaded histogram),  the background (solid  histogram) and the
 data (points with error bars) for $\sqrt{s}$ from
 204 to 209 GeV for the b-jet energy (left) and for the b-tag variable (right).
}
\label{distr1}
\end{figure}
 The distributions of the b-jet energy and  of the b-tag variable 
 are shown in Fig.~\ref{distr1}.

 The signal selection efficiencies and
 the expected backgrounds at the various centre-of-mass energies 
 are reported in Table~\ref{tabellone}. The efficiencies are given for 
 a top mass of 174~$\gevcc$ and ${\rm BR(t\to b W)}=1$.

 The main sources of systematic uncertainties on the 
 expected background and the signal efficiencies
 have been assessed as in Ref.~\cite{tc1}.
 In order to increase the statistical precision,
 the data collected by ALEPH at lower centre-of-mass
 energies in the years 1998 and 1999~\cite{tc1} have been included
 in these systematic checks, except for the one on the b-tag variable
 which can vary from year to year. 
 The results of these systematic studies are reported in Table~\ref{syst}.

\section{Results}
\label{results}
 The number of candidate events are given in Table~\ref{tabellone} 
 for the various centre-of-mass energies. 
 In total 24 candidates were observed in the data in agreement
 with 20.1 events expected from SM backgrounds.
 Upper limits on the signal cross section have been 
 derived at 95$\%$ CL for the different channels and centre-of-mass energies
 (Table~\ref{tabellone}).

\begin{table}[p]
\caption{\small \label{tabellone} Performance and results of the analysis.
 At each centre-of-mass energy the numbers of expected background events 
($N^{\rm bkg}$), of observed candidates ($N^{\rm obs}$), 
the signal efficiency $\varepsilon$ computed with respect to all W decays, 
and the expected and measured 95\%~CL upper limits
on the single top production cross section
 ($\sigma_{95}^{\rm exp}$ and $\sigma_{95}^{\rm meas}$) 
are reported for both the leptonic and hadronic W decays. 
 Systematic uncertainties are 
 not included in these cross section upper limits.
 In the last row the measured 95\%~CL upper limits on the single top
 production cross section ($\sigma_{95}^{\rm comb}$), 
 obtained by combining the leptonic and  hadronic channels 
 and including the systematic 
 uncertainties on background and on the signal efficiencies are given.}
\begin{center}
\begin{tabular}{|c||c|c||c|c||c|c||c|c||c|c|}
\hline
$<$\roots$>$~(GeV) & \multicolumn{2}{|c||}{203.8}&\multicolumn{2}{|c||}{205.0}&\multicolumn{2}{|c||}{206.3}&\multicolumn{2}{|c||}{206.6} &\multicolumn{2}{|c|}{208.0} \\
\hline
\rule{0pt}{4.5mm}
${\cal L}$~(pb$^{-1}$) & \multicolumn{2}{|c||}{8.3}&\multicolumn{2}{|c||}{71.6}&\multicolumn{2}{|c||}{65.6}&\multicolumn{2}{|c||}{61.0} &\multicolumn{2}{|c|}{7.3} \\
\hline
 & lept. & hadr.  & lept. & hadr.  & lept. & hadr.  & lept. & hadr.  & lept. & hadr. \\
\hline 
          &  &  &      &       &       &      &      &     &      &      \\[-.9pc]
$N^{\rm bkg}_{\rm WW}$  &0.1 &0.4 &0.6 &3.4  &0.6  &3.1  &0.5  &2.9 &0.1 &0.4 \\
$N^{\rm bkg}_{\rm 4f}$  & -  &0.1 &0.1 &0.7  &0.03 &0.7  &0.05 &0.6 &-   &0.1 \\
$N^{\rm bkg}_{\rm qq}$  & -  &0.2 & -  &1.9  &-    &1.7  & -   &1.6 &-   &0.2 \\
          &  &  &      &       &       &      &      &     &      &      \\[-.9pc]
\hline
          &  &  &      &       &       &      &      &     &      &      \\[-.9pc]
$N^{\rm bkg}_{\rm tot}$ &0.1 &0.7 &0.7 &6.0  &0.63 &5.5  &0.55 &5.1&0.1 &0.7 \\
$N^{\rm obs}$           & 0  & 1  & 2  &6    &0    &7    &0    &7  &0   &1 \\
\hline
          &  &  &      &       &       &      &      &     &      &      \\[-.9pc]
$\varepsilon$ (\%)     & 3.4  & 13.8   & 3.3  & 13.7 & 3.3 & 13.7  & 3.2  & 13.5 & 3.3 & 13.2 \\
$\sigma_{95}^{\rm exp} $~(pb)
                       & 11.2 & 3.4    & 1.65 & 0.71 & 1.77& 0.75  & 1.91 & 0.80 & 13.3& 4.10 \\
$\sigma_{95}^{\rm meas}$~(pb)
                       & 10.6 & 3.7    & 2.38 & 0.69 & 1.38& 0.89  & 1.53 & 1.01 & 12.6& 4.47  \\
\hline
$\sigma_{95}^{\rm comb}$~(pb) & \multicolumn{2}{|c||}{ 2.94  }&\multicolumn{2}{|c||}{ 0.68 }&\multicolumn{2}{|c||}{ 0.68 }&\multicolumn{2}{|c||}{ 0.78 } &\multicolumn{2}{|c|}{ 4.32 } \\
\hline
\end{tabular}
\end{center}
\end{table}

\begin{table} [p]
\caption{\small 
\label{syst} Relative systematic uncertainties on the background and signal
 efficiency for each selection variable~\cite{tc1},
 determined by applying one cut at a time in data and Monte Carlo.}
\begin{center}
\begin{tabular}{|c|c||c|c|}
\hline
\multicolumn{2}{|c||}{Leptonic W}&\multicolumn{2}{c|}{Hadronic W}\\
Variable & $\frac{\varepsilon_{\rm data}-\varepsilon_{\rm MC}}{\varepsilon_{\rm MC}}$ (\%) 
& Variable &  $\frac{\varepsilon_{\rm data}-\varepsilon_{\rm MC}}{\varepsilon_{\rm MC}}$ (\%)\\[1mm]
\hline
Lepton ID               & $3.8\pm2.9$         & $P_{\rm qq\rm qq}/E_{\rm qq}$  & $1.4\pm1.5$    \\
$m_{\rm qq}$       & $1.0\pm3.4$         & $m_{\rm qq}$        & $-1.1\pm1.5$   \\
$E$(b jet)$_{\rm t cm}$ & $0.5\pm1.1$    & $E$(b jet)               & $0.1\pm0.9$    \\ 
$m_{{\rm {\ell}} \nu}$  & $1.9\pm2.1$         & Thrust                   & $0.1\pm0.7$    \\
$\mt$                   & $-0.8\pm1.4$        & $E_{\rm tot}$            & $-0.1\pm0.4$   \\
Highest jet b tag       & $5.1\pm4.7$         & Highest jet b tag        & $1.3\pm4.0$    \\
\hline 
\end{tabular}
\end{center}
\end{table}

 The negative result of these searches are translated into limits on the
 top quark couplings  $\kZ$ and $\kg$~\cite{pecceia}. 
 The data collected by ALEPH at lower centre-of-mass
 energies in 1998 and 1999~\cite{tc1} are also included to derive
 the exclusion region in the ($\kZ$, $\kg$) plane and in the 
 [BR($\rm{t \to Z c/u}$), BR($\rm{t \to \gamma c/u}$)] plane.
 The likelihood ratio method~\cite{like1} has
 been used in the computation of the excluded regions.
 The signal selection
 efficiency and the background expectation
 were conservatively reduced by their systematic uncertainties.

\begin{figure}[p]
 \begin{center}
 \mbox{\epsfig{file=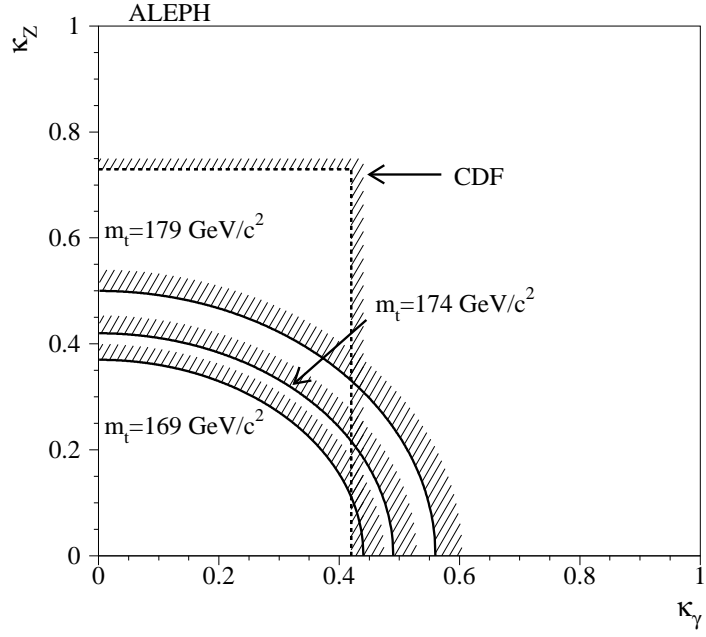,width=10cm,height=10cm}}
 \end{center}
 \caption[]
 {\small Exclusion curves at 95\%~CL in the ($\kZ$, $\kg$) plane 
 for $\mt =  169,~174,~179~  \gevcc$ (full lines).
 The region excluded by CDF  is also shown (dotted line).}
 \label{kklimits}
 \end{figure}

\begin{figure}[p]
 \begin{center}
 \mbox{\epsfig{file=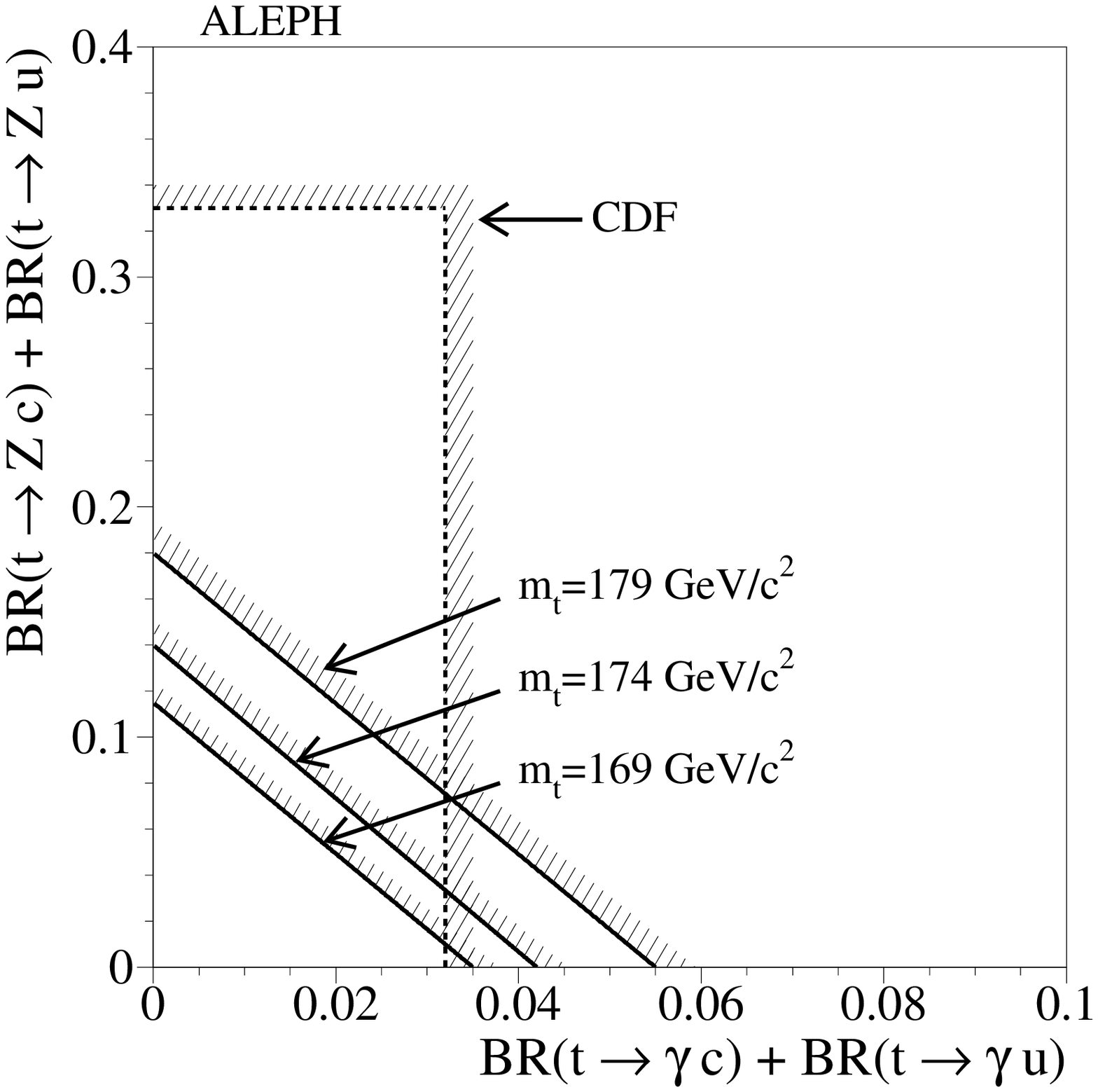,width=10cm,height=10cm}}
 \end{center}
 \caption[]
 {\small Exclusion curves at 95\%~CL in the 
 [BR($\rm{t \to Z c/u}$), BR($\rm{t \to \gamma c/u}$)]
 plane for $\mt = 169,~174,~179~  \gevcc$ (full lines).
  The region excluded by CDF  is also shown (dotted line).}
 \label{bblimits}
 \end{figure}

 The region of the ($\kZ$,$\kg$) 
 plane excluded at 95\%~CL is shown in Fig.~\ref{kklimits}. 
 The excluded region in the  
 [BR($\rm{t \to Z c/u}$),BR($\rm{t \to \gamma c/u}$)] plane
 is shown in Fig.~\ref{bblimits}.
 The limits are given for different choices of the top mass. 
 The exclusion curves include the reduction in BR($\rm t \to bW$),
 computed as a function of $\kZ$ and $\kg$,
 due to possible FCNC decays of the top.
 The limits obtained by CDF from a search for 
 the decays ${\rm t \to Z c, Z u }$ and
 ${\rm t \to \gamma c, \gamma u }$~\cite{cdf_tc} are also shown.

A 95\% CL upper limit of 0.42 for the anomalous coupling $|\kZ|$ 
is obtained assuming $\mt=174$~$\gevcc$ and $\kg=0$. 
This exclusion translates into the  branching ratio limit 
${\rm BR(t\to Zc)+BR(t\to Zu)}<14 \%$.

\section{Conclusions}
\label{Conclusions}
Single top production via flavour changing neutral currents
 has been searched for in 214~pb$^{-1}$ of data collected by
 ALEPH at centre-of-mass energies between 204 and 209 GeV. 
In total, 24 events have been selected in the data in agreement  
with 20.1 expected from Standard Model backgrounds. 
Upper limits at 95\%~CL on  single top production cross sections at  
$\roots=204$--209~GeV have been derived. 

The combination with the  data  collected in 1998 and 1999 yields 
 a 95\%~CL upper limit on the FCNC coupling 
for Z exchange of $|\kZ| < 0.42$, for $\mt=174$~$\gevcc$ and  $\kg=0$. 
It corresponds to a branching ratio limit of
${\rm BR(t\to Zc)+BR(t\to Zu)}<14 \%$.
 This result updates the previous ALEPH measurements at lower centre-of-mass 
 energies~\cite{tc1},
 and is in agreement with recent results from OPAL~\cite{opal}.

\section*{Acknowledgements}
We thank our colleagues from the CERN accelerator divisions for the successful running  
of LEP at high energy.
We are indebted to the engineers and technicians in all our institutions for their contribution
to the continuing good performance of ALEPH. Those of us from non-member states thank CERN
 for its hospitality.

\end{document}